# Data-driven kinetic energy density fitting for orbital-free DFT: linear vs Gaussian process regression

Sergei Manzhos[a,1] and Pavlo Golub[b]

[a] Centre Énergie Matériaux Télécommunications, Institut National de la Recherche Scientifique, 1650 boulevard Lionel-Boulet, Varennes QC J3X 1S2 Canada.

[b] J. Heyrovsky Institute of Physical Chemistry, Department of Theoretical Chemistry, Dolejškova 2155/3, 182 23 Prague 8, Czech Republic

**Abstract**

We study the dependence of kinetic energy densities (KED) on density-dependent variables that have been suggested in previous works on kinetic energy functionals (KEF) for orbital-free DFT (OF-DFT). We focus on the role of data distribution and on data and regressor selection. We compare unweighted and weighted linear and Gaussian process regressions of KED for light metals and a semiconductor. We find that good quality linear regression resulting in good energy-volume dependence is possible over density-dependent variables suggested in previous literature. This is achieved with weighted fitting based on KED histogram. With Gaussian process regressions, excellent KED fit quality well exceeding that of linear regressions is obtained as well as a good energy-volume dependence which was somewhat better than that of best linear regressions. We find that while the use of the effective potential as a descriptor improves linear KED fitting, it does not improve the quality of the energy-volume dependence with linear regressions but substantially improves it with Gaussian process regression. Gaussian process regression is also able to perform well without data weighting.

## 1 Introduction

The development of orbital-free density functional theory (OF-DFT)[1,2] faces two major hurdles: (1) the accuracy of kinetic energy functional (KEF) approximations and (2) the absence of quality local pseudopotentials (LPP) for most elements of the periodic table which would allow KEF approximations on smoother (pseudo)valence density. Good LPP exist for several light metals and semiconductors,[3-10] and existing KEFs already provide an accuracy sufficient (i.e. similar to Kohn-Sham (KS) DFT)[11] for large-scale modelling of light metals in applications.[6,7,12-14] For semiconductors, however, the accuracy is still problematic for use in applications[15,16] even when decent LPP exist for some semiconductors.[7] Similar to developments in KS-DFT, where a range of functionals have been developed and are used for different classes of materials, OF-DFT can be useful in applications as long as there are KEFs with portability limited to certain types or groups of materials. Nevertheless, it is important to have KEFs capable of handling both metals and semiconductors, as many practically important types of systems involve interactions between metals and semiconductors or phase transitions between metallic and semiconducting phases.[16-20] In this work, we therefore consider two types test systems: solid Al, Mg, and Si, i.e. two metallic and one semiconductor. We therefore ensure a degree of generalization in the presented models, although the purpose of this work is not to achieve wide generalization.

---

[1] Author to whom correspondence should be addressed. E-mail: sergei.manzhos@emt.inrs.ca . Tel.: +1-514-228-6841.

The promise of OF-DFT lies in enabling routine large-scale ab initio calculations. Even without access to bandstructure, OF-DFT can bring ab initio accuracy to types of simulations currently dominated by force fields due to system sizes that make near-cubically scaling KS-DFT simulations prohibitive or unwieldy. This promises better-than-force field accuracy as well as access to some electronic properties including electron density itself, which allow performing analyses which are in principle impossible with force fields.[21,22] Linear-scaling KS-DFT approaches have been proposed[23-25] but remain costly (large prefactors) and introduce additional approximations. OF-DFT inherently can have near-linear scaling, but not all KEF models are linear-scaling. Notably, non-local KEFs[15,26-30] are not generally linear-scaling, due to the need to compute the integral representing the non-local part $E_{NL}[\rho(r)]$ of the KEF. Typically, such functionals represent the kinetic energy (KE) as

$$E_{kin}[\rho(r)] = E_{TF}[\rho(r)] + E_{vW}[\rho(r)] + E_{NL}[\rho(r)]$$

where $\rho(r)$ is the space-dependent electron density, $E_{TF}[\rho(r)] = \frac{3}{10}(3\pi)^{2/3} \int \rho^{5/3}(r)dr$ is the Thomas-Fermi (TF) model[31] and $E_{vW}[\rho(r)] = \frac{1}{8}\int \frac{|\nabla\rho(r)|^2}{\rho(r)} dr$ is the von Weizaecker (vW) model[32] (atomic units are used unless indicated otherwise). The non-local term is usually expressed as

$$E_{NL}[\rho(r)] = \iint \rho^\alpha(r)\omega(r,r')\rho^\gamma(r')drdr'$$

which in general results in a quadratic to $n\,log(n)$ scaling.

Semi-local KEFs, on the other hand, are inherently near-linear scaling. Semi-local KEFs model the kinetic energy density KED and are therefore KED functionals (KEDF). While it is not strictly necessary to reproduce KS KED with a KEDF to obtain an accurate KE, typically KEDFs are built to reproduce either the KS KED based on the KS equation:

$$\tau_{KS}(r) = -\frac{1}{2}\sum_i \psi_i(r)\Delta\psi_i(r)$$

or its positively definite version

$$\tau_+(r) = \frac{1}{2}\sum_i |\nabla\psi_i(r)|^2 = \tau_{KS}(r) + \frac{1}{4}\Delta\rho(r)$$

Here are $\psi_i(r)$ are Kohn-Sham orbitals, and the sum here and below is over all occupied orbitals (without loss of generality, we ignore spin and partial occupancy). Both $\tau_{KS}$ and $\tau_+$ will be referred to as KS KED as both depend on the orbitals. The Laplacian of the density integrating to zero, both $\tau_{KS}$ and $\tau_+$ result in the same $E_{kin}$. One can define TF and vW KEDs as, respectively, $\tau_{TF}(r) = \frac{3}{10}(3\pi^2)^{2/3}\rho^{5/3}(r)$ and $\tau_{vW}(r) = \frac{1}{8}\frac{|\nabla\rho(r)|^2}{\rho(r)}$. The TF KEDF is a local functional and vW KEDF a semi-local functional due to its dependence on the derivatives on the density. In general, one seeks to build a KED model as a function of different density-dependent variables including derivatives and powers of the density of different orders:

$$\tau[\rho(r), \nabla\rho(r), \Delta\rho(r), ...] \approx \tau_{KS/+}(r)$$

Recently, it was shown that Laplacian-level semi-local KEDFs could accurately model some metallic as well as covalent systems. This included functionals built by rational, intuition driven designs as well as with the help of machine-learning (ML).[33] For example, Constantin et al.[34] proposed a semi-local KEDF (PGSL) that showed attractive accuracy for both metals and semiconductors. Their functional has the form

$$\tau_{PGSL}[\rho(r)] = \tau_{TF}\left(\exp\left(-\frac{40}{27}p\right) + \beta q^2\right)$$

with a recommended $\beta = 0.25$. Here $p = \frac{|\nabla\rho|^2}{4(3\pi^2)^{2/3}\rho^{8/3}}$ is the scaled squared gradient and $q = \frac{\Delta\rho}{4(3\pi^2)^{2/3}\rho^{5/3}}$ is the scaled Laplacian. The scaling helps satisfy the so-called exact conditions.[35] Luo et al.[36] proposed the LKT functional of the form

$$\tau_{LKT}[\rho(r)] = \tau_{TF}\left(\frac{1}{\cosh(1.3p^{1/2})} + \frac{5}{3}p\right)$$

which performed well on several metals and semiconductors. Golub and Manzhos[37] performed neural network (NN) fitting of KED as function of the terms of the 4$^{th}$ order gradient expansion,[38] $\tau_{GE4} = \tau_{GE2} + \tau_{TF}\left(\frac{8}{81}q^2 - \frac{1}{9}pq + \frac{8}{243}p^2\right)$, where $\tau_{GE2} = \tau_{TF}\left(1 + \frac{5}{27}p + \frac{20}{9}q\right)$ is the 2$^{nd}$ order gradient expansion (which includes the vW KED in the second term[39]). They were able to achieve good fits to KS KEDs for metallic as well as covalently bonded systems. The Nakai group fitted KEDs of a range of molecular systems with NNs.[40,41] They used up to 3$^{rd}$ order (unscaled) gradients of the density. With the inclusion of the information on atomic positions into the variable set (in the form of distances to the atoms) they were able to obtain smooth potential energy curves.[41] Inclusion of such data introduces non-local information but goes beyond using density-dependent variables. Another way to introduce such information is to use the Hartree or KS effective potential; this would have the advantage of being a density-dependent variable. *One question we address in the present work in therefore what is a good set of density-dependent variables to build a KEDF? Specifically, what is the utility of the effective potential as variable?*

Golub and Manzhos[37] noted that most of KED variance could be captured by a linear fit. The non-linear NN size (number of neurons) was relatively small. While for a combined NN fit of KEDs of several materials, they had to use a deep NN, small single-hidden layer NNs were able to accurately reproduce the KED of a given material. A NN is a linear combination of simple non-linear functions. That non-linearity could potentially be subsumed in the definitions of the density-dependent variables. For example, the promising PGSL functional can be viewed as a linear combination of the set of variables $\left\{\tau_{TF}, \ \tau_{TF}\exp\left(-\frac{40}{27}p\right), \ q^2\right\}$. *Another question we address in the present work is therefore whether with a proper choice of density dependent variables a linear fit could be sufficient to produce a useful KEDF; in particular, whether energy-volume curves can be reproduced which are key to structural optimization.*

Data distribution and its handling appear to be critical issues in KED fitting. Golub and Manzhos emphasized the very uneven nature of KED distributions.[37] They were able to obtain good quality fits after performing data selection for NN fitting (a fraction of the total data was used for fitting) in a way that made the distribution of fitting data less uneven than that of the raw KED data. The reason why single-hidden layer NNs could produce excellent KED fits for individual materials without overfitting while a deep NN was needed to fit the KEDs of several materials simultaneously was also related to the data distribution.[37] *The third question we address in this work is how can one act on the data distribution to improve KEDF fitting?* We therefore compare KED regressions without and with data weighting based on KED distributions.

Machine learning (ML) of KEFs has been gaining momentum, with multiple works showing promising results for synthetic and recently real-life systems.[33,37,40,41-48] Comparisons among approaches and between ML and non-ML methods are, however, scarce. In this work, we also use Gaussian process regression (GPR) to learn the KED. GPR is similar in the form of representation to the kernel ridge regression (KRR) method[49] which has been used in several works to fit the KED;[43-45] in GPR, the terms of the representation are found with the Bayesian approach.[50,51] GPR has been shown to be superior to neural networks in other applications.[52] *Another purpose of the present work is to compare the performance of linear and Gaussian process regression, and in particular their behavior with respect to variables choices and data weighting.*

We consider both the KED fit quality and total energy quality of the models, specifically the ability to reproduce energy-volume dependence which is critical for structural optimization.

## 2    Methods

### 2.1   Kohn-Sham DFT simulations

KS-DFT calculations were performed in Abinit.[53,54] The PBE exchange-correlation functional[55] was used. The plane-wave cutoff for the expansion of the orbitals was set to 500 eV, and we confirmed that no changes were observed when increasing it to 1000 eV. Bulk-derived, real space local pseudopotentials from Carter's group were used.[7] Energies were converged to $1\times10^{-7}$ *a.u.*, and structures were optimized until all force components were below $1\times10^{-4}$ *a.u.* Conventional unit cells were used for face-centered cubic Al, cubic diamond Si, and a hexagonal closed packed cell for Mg. The Brillouin zone was sampled with 8×8×8, 6×6×6, and 10×10×8 *k*-points, respectively. Optimal lattice constant were 4.047 Å for Al and 5.470 Å for Si while for Mg $a$ = 3.18 Å, $c$ = 5.25 Å, in good agreement with literature.[56] Densities, kinetic energy densities ($\tau_+$), gradients and Laplacians of the density, as well as Kohn-Sham effective potentials on the entire Fourier grid (the real space grid equivalent to the plane wave cutoff energy) were output from Abinit at the equilibrium geometry as well as at isotropically compressed or expanded lattice constants by 1-δ and 1+ δ (volume changes by $(1\pm\delta)^3$), respectively, with δ=0.05.

### 2.2   Regressions

From the density, its gradients, its Laplacian, and the KS effective potential $V_{eff}(\boldsymbol{r})$ output by Abinit, the following density dependent variables were formed: $\tau_{TF}, \tau_{vW}, \tau_{TF}p, \tau_{TF}q, \tau_{TF}p^2, \tau_{TF}pq, \tau_{TF}q^2, \tau_{TF}\exp\left(-\frac{40}{27}p\right), \frac{\tau_{TF}}{\cosh(1.3p^{1/2})}, \rho V_{eff}$ . We also attempted fits of KED as functions of fixed combinations of some of these variables such as $\tau_{LKT}$, $\tau_{PGSL}$, $\tau_{GE2}$, and $\tau_{GE4}$ corresponding to LKT, PGSL functionals and 2$^{nd}$ and 4$^{th}$ order gradient expansions, respectively.

Linear regressions were performed in Octave[57] in a home-made code on the entire Fourier grid. Non-weighted and weighted regression (using *LinearRegression* function) were performed. Gaussian process regressions were performed in Octave in a home-made code using the STK module. The reader is referred to literature for the description of GPR.[50,51] For GPR, data were scaled to the range [0, 1] and an isotropic Matern3/2 covariance function was used. No significant advantage was found using anisotropic kernels or squared exponential, exponential, or Matern5/2 kernels. The length parameter of the covariance function was chosen to be 0.5; values differing on the order of 0.1 were not significantly affecting the results. Gaussian noise amplitude was chosen on the order of $1\times10^{-2}$. In GPR, only a small fraction of the data was used for model training while all data were used to report regression quality and energy-volume behavior.

In weighted regressions, the histogram of $\tau_+$ values $h(\tau_+)$ was used to define the weights. The weights were assigned as $w(\tau_+) = h^{-1}(\tau_+)$. A smooth $h(\tau_+)$ was obtained by cubic spline interpolation of a histogram computed using 200 bins. The weights were used as is in linear regressions. In Gaussian process regressions, they were used to drive data selection for the training dataset: points were accepted into the training set for which $w'(\tau_+) > rand$, where $w'$ is the function $w$ normalized to a maximum of 1 and $rand$ is a random number. With this selection scheme sometimes very few points are selected from a total set of close to 590,000 points; in that case the selection is repeated (with different random numbers) and the points are accumulated until a desired number of training points $N_{train}$ is obtained. We used $N_{train}$ = 2,000 and 5,000 training points. The number of training points did not exceed 1% (!) of the total point set and overfitting is therefore not an issue; computed Pearson $R^2$ coefficients, *rmse* (root mean squared error), and *B'* values (defined in section 2.3) reflect therefore predictive power of GPR. When using unweighted data, we randomly and uniformly selected the training points from the total set. In both linear regressions and GPR, the Pearson coefficients, *rmse*, and *B'* values are computed on all data points. To avoid bias associated with a specific selection of $N_{train}$ points, we averaged over 5 runs, effectively forming a GPR committee (see e.g. Ref. 58 on the use of committees). This effectively expands the ratio of training points but only to about 4% of the total dataset. Unless stated otherwise, results are presented for fits of a dataset combining data from all three materials (Al, Mg, and Si) and at three geometries (i.e. equilibrium, compressed, and expanded).

*2.3   Quality of total energy prediction*

In the limit of infinite accuracy of KED fitting, it would result in accurate kinetic and total energies for different geometries, and thereby provide a functional accurate for structure optimization. With finite accuracy, this is not generally the case. The quality of the regression over KED values as expressed by the Pearson $R^2$ coefficient or error measures such as *rmse* may not be reflective of the quality of KE prediction. This has to do with the fact that dependence of energy on geometry is due to small relative changes in the energy, and unless KED errors are extremely small, a degree of error cancellation is desired. We therefore test the performance of different regressions when computing total energies at different geometries. We define the following quantity:

$$B' = V_0^2 \frac{d^2E}{dV^2} \approx \frac{E(V_0 - \Delta V) - E(V_0) + E(V_0 + \Delta V)}{(\Delta V/V_0)^2}$$

which is related to the curvature of the energy-volume dependence (and therefore to bulk modulus) but is easier to estimate based on our data. The energies $E$ here are obtained by replacing the KS kinetic energy from Abinit with an integrated KED from the model. Here $V_0$ is the equilibrium volume of the simulation cell and $\Delta V/V_0 = (1 + \delta)^3 - 1$. As this quantity is related to the curvature of the energy-volume curve at the equilibrium geometry, it is indicative of the capacity of the KED approximation to find the optimal structure. We do not address in this work issues related to self-consistent density optimization.

## 3 Results

### 3.1 Data and variables distributions

Isosurfaces of the density, KED, the Laplacian of the density, and the KS effective potential of Al and Si at the equilibrium geometry are illustrated in **Figure 1**. The more uniform density of Al and Mg indicative of metallic bonding and more directional density pattern of Si indicative of covalent bonding can be seen in these plots. The distributions of the KED as well as of the density-dependent variables considered here are shown in **Figure 2** to **Figure 5** for, respectively, Al, Mg, and Si as well as for a combined dataset for Al, Mg, and Si, at all geometries. The distributions are highly uneven. Specifically, the scaled gradient and Laplacian ($p$ and $q$), and even more so their squares and products, are extremely unevenly distributed. TF KED distribution is relatively less uneven compared to $p$- and $q$- based variables but remains very non-uniform. This complicates any KED fitting and makes it necessary to tackle these distributions. Moreover, the relative degree of non-uniformity of the distributions is system-dependent, e.g. with Al, $\tau_{KS}$ appears to be less uneven compared to $\tau_+$, but the opposite is the case with Si. To correctly reproduce the KE, it is important to correctly account for the bulk of the data as well as the thinly sampled tails of the distributions. Including different geometries in the data further complicates the distributions. They remain extremely uneven, but the number of features increases. The KED fitting in the subsequent section will be performed on the combined dataset of the three materials while the quality of the KE prediction by the models will be tested for Al, Mg, and Si individually.

*3.2 Linear KED fitting*

We first present the quality of the KED and of the energy-volume dependence (*B'*) obtained with the common TF, vW, and TF+vW approximations, the 4$^{th}$ order gradient expansion, and the recent semi-local KEDFs LKT and PGSL. These are shown in **Figure 6** as correlation plots between the KS KED and the respective models. The correlations are very weak, and do not appear to be necessarily better for LKT and PGSL than for the primitive TF+vW model. We will show below, however, that this does not prevent some of these approximations to achieve reasonable KE as a function of volume. As explained above, while a perfect KED match would guarantee an accurate KE, in an approximate model, to obtain the best KE, the best possible match between the KS KED and the model is not necessarily what one should strive for even when fitting a KEDF.

We now present the results of regressions of the KED over selected combinations of the density-dependent variables. These combinations are:

Model1: $\{\tau_{TF}, \tau_{TF}p \mid \rho V_{eff}\}$

Model2: $\{\tau_{TF}, \tau_{TF}p, \tau_{TF}q \mid \rho V_{eff}\}$

Model3: $\{\tau_{TF}, \tau_{TF}p, \tau_{TF}q, \tau_{TF}p^2, \tau_{TF}pq, \tau_{TF}q^2 \mid \rho V_{eff}\}$

Model4: $\{\text{LKT, PGSL}, \tau_{TF}q \mid \rho V_{eff}\}$

Model5: $\{\tau_{TF}, \tau_{TF}p, \tau_{TF}q, \tau_{TF}q^2, \tau_{TF}\exp\left(-\frac{40}{27}p\right) \mid \rho V_{eff}\}$

Model6: $\{\tau_{TF}, \tau_{TF}p, \tau_{TF}q, \tau_{TF}q^2, \frac{\tau_{TF}}{\cosh(1.3p^{1/2})} \mid \rho V_{eff}\}$

where $\mid \rho V_{eff}$ in each model denotes that the effective potential-dependent term may or may not be included; we report below regression results for both cases. The regression plots for weighted and unweighted linear regressions when including or not the $\rho V_{eff}$ term are shown in to **Figure 7** to **Figure 10**. In **Table 1**, we report correlation coefficients and KED *rmse* values obtained with these models as well as the *B'* values. It follows from these results that:

- Among the known functionals chosen here for comparison, all but vW are able to reproduce a minimum in the energy-volume dependence (positive *B'*) for all Al, Mg, and Si. The quality of this minimum is quite low with the TF-vW based models and improved with higher order (Laplacian level) models. The 4$^{th}$ order gradient expansion beats the selected recent KEDFs for the metals Al, Mg as well as for the semiconductor Si.

- It is possible to obtain with a linear regression model KED fit quality (expressed by high $R^2$ and low *rmse*) far exceeding that obtained with simple TF and vW approximations (or their combination), the 4$^{th}$ order gradient expansion, as well as that of the chosen recent semi-local KEDFs. However, this does not signify significant improvement in the quality of prediction of kinetic energy as a function of geometry. Using all terms entering GE4 as density-dependent variables provides the best results. Adding terms from LKT or PGSL functionals (Model5 and Model6) does not seem to bring additional advantages.

- Weighting the fit with the distribution of KED values can substantially improve linear models, especially for Si, where the quality of *B'* is low in an unweighted fit. Any improvement is less clear for the metals. Model3, i.e. using all components of the 4$^{th}$ order gradient expansion as variables, appears to perform best and beats the gradient expansion in that it is able to reproduce *B'* well simultaneously for Al, Mg, and for Si. This supports the strategy of using terms of the 4$^{th}$ order expansion as density-dependent variables to build more involved models.[37]

- Comparison of the results with and without the $\rho V_{eff}$ as variable is most interesting: while inclusion of this term significantly improves fit quality, it also significantly worsens the quality of energy-volume dependence prediction, so much so that for Si with most models *B'* turns negative, i.e. the model does not have a minimum near the equilibrium geometry.

The last point is important in that it points to limits of linear models. The kinetic energy can be expressed as a sum of Kohn-Sham eigenvalues $\sum_i \epsilon_i$ and a functional of $\rho V_{eff}$. The KED can be expressed as a sum of a term proportional to $\rho(r)V_{eff}(r)$ and $\sum_i |\psi_i(r)|^2 \epsilon_i$. Indeed, this follows directly from multiplying the Kohn-Sham equation for *i*-th orbital on the left by $\psi_i(r)$ (or by its complex conjugate if the orbitals are complex; we work with real orbitals) and summing over *i*:

$$-\frac{1}{2}\Delta\psi_i(r) + V_{eff}(r)\psi_i(r) = \epsilon_i\psi_i(r) \quad |\times \psi_i(r), \sum_i$$

$$-\frac{1}{2}\sum_i \psi_i(r)\Delta\psi_i(r) + V_{eff}(r)\sum_i |\psi_i(r)|^2 = \sum_i \epsilon_i|\psi_i(r)|^2$$

$$\tau_{KS}(r) + V_{eff}(r)\rho(r) = \sum_i \epsilon_i|\psi_i(r)|^2$$

**Table 1.** Correlation coefficients $R^2$, room mean square error of the KED *rmse*, and proxies of curvatures of the energy-volume dependence $B'$ for Al, Mg, and Si following from different regression models. The best values among the models of similar types are highlighted in bold, and qualitatively wrong values in red italic font. The best overall models in each set (unweighted/weighted, with/without $\rho V_{eff}$) is highlighted in green. For GPR, results with 2,000 and 5,000 train points per GPR are given in top and bottom rows, respectively. The GPR used the variables of Model3.

| | model | $R^2$ | *rmse* | $B'$(Al) | $B'$(Mg) | $B'$(Si) |
|---|---|---|---|---|---|---|
| | (reference $B'$:) | | | 1.0595 | 0.36456 | 3.0917 |
| | TF | 0.49463 | 0.00605 | 0.72863 | 0.28207 | 1.9249 |
| | vW | 0.60117 | 0.00772 | *-0.3284* | 0.05296 | *-2.1211* |
| | TF+vW | 0.69794 | 0.00564 | 2.1688 | 0.50861 | **3.0799** |
| | GE4 | 0.67916 | **0.00395** | **0.9718** | **0.32193** | 2.1455 |
| | LKT | 0.64363 | 0.00546 | 1.5687 | 0.41646 | 2.6888 |
| | PGSL | 0.64796 | 0.00554 | 1.4867 | 0.41329 | 2.8177 |
| unweighted | 1 | 0.8725 | 0.00265 | 2.0615 | 0.45902 | 1.3288 |
| | 2 | 0.87158 | 0.00245 | 1.5856 | 0.39765 | 1.6336 |
| | 3 | 0.87066 | **0.00244** | 1.5616 | **0.39257** | 1.7362 |
| | 4 | 0.84904 | 0.00254 | **1.1251** | 0.33592 | **1.8445** |
| | 5 | 0.87062 | 0.00245 | 1.5943 | 0.39819 | 1.618 |
| | 6 | 0.87063 | 0.00245 | 1.5936 | 0.39819 | 1.6211 |
| | GPR | 0.98526 | 0.00074 | 0.91082 | 0.18081 | *-3.6066* |
| | | 0.98982 | 0.00061 | 1.0950 | 0.23882 | *-3.7413* |
| weighted | 1 | 0.8423 | **0.00293** | 1.9423 | 0.44852 | 1.6535 |
| | 2 | 0.82776 | 0.00323 | 1.7674 | 0.43772 | 2.3631 |
| | 3 | 0.77746 | 0.00423 | **1.2910** | **0.3896** | 3.3940 |
| | 4 | 0.81544 | 0.00338 | 1.5353 | 0.4100 | 2.6157 |
| | 5 | 0.79544 | 0.00398 | 1.4655 | 0.40982 | **3.2077** |
| | 6 | 0.79525 | 0.00398 | 1.4668 | 0.4102 | 3.2147 |
| | GPR | 0.98306 | 0.00082 | 1.0191 | *-0.0104* | 0.36691 |
| | | 0.98908 | 0.00065 | 0.36751 | 0.0337 | 2.3221 |
| unweighted, with $\rho V_{eff}$ | 1 | 0.88828 | 0.00252 | 2.0811 | **0.4640** | 0.66143 |
| | 2 | 0.9897 | 0.00066 | **0.40312** | 0.25607 | *-0.5196* |
| | 3 | 0.99114 | 0.00062 | 0.3381 | 0.24459 | *-0.2546* |
| | 4 | 0.98984 | 0.0007 | 0.26794 | 0.23813 | *-0.4932* |
| | 5 | 0.99156 | **0.0006** | 0.29179 | 0.24305 | *-0.463* |
| | 6 | 0.99154 | 0.00061 | 0.29094 | 0.24268 | *-0.4701* |
| | GPR | 0.99876 | 0.00021 | 0.91358 | 0.44368 | 2.0051 |
| | | 0.99972 | **0.00010** | 1.0147 | 0.42109 | 2.9808 |
| weighted, with $\rho V_{eff}$ | 1 | 0.76359 | 0.00404 | 2.1059 | 0.48085 | **3.0597** |
| | 2 | 0.98537 | **0.00077** | 0.55256 | 0.27843 | *-0.5205* |
| | 3 | 0.9803 | 0.00113 | 0.49629 | 0.2738 | 0.37026 |
| | 4 | 0.97979 | 0.00114 | **0.6414** | **0.29997** | 0.2954 |
| | 5 | 0.98538 | 0.00092 | 0.61118 | 0.2895 | 0.02404 |
| | 6 | 0.98519 | 0.00092 | 0.61409 | 0.28986 | 0.03094 |
| | GPR | 0.99958 | 0.00012 | 0.97374 | 0.55362 | 3.4288 |
| | | 0.99989 | 0.00006 | 0.98999 | 0.46950 | 3.3481 |

It is not surprising that the inclusion of the $\rho V_{eff}$ term substantially improves the linear fit as it captures a significant fraction of the variance. Indeed, for example, for Model3 the coefficients at the variables are {0.177, -0.118, 2.767, 0.339, -0.201, -0.007, -0.647}, the last value being the weight of the $\rho V_{eff}$ term (we note that all linear coefficients remained well defined, i.e. their values are higher that their uncertainties). However, this simply shifts the problem to fitting the orbital-weighted band energy ($\sum_i |\psi_i(\boldsymbol{r})|^2 \epsilon_i$) for which linear models appear to be not powerful enough. In the next section, therefore, we explore a type of non-linear regression using a machine learning method (GPR).

*3.3  Non-linear models with Gaussian process regression*

The results of the previous section suggest that all variables in the set $\{\tau_{TF},\ \tau_{TF}p,\ \tau_{TF}q,\ \tau_{TF}p^2,\ \tau_{TF}pq,\ \tau_{TF}q^2\ |\ \rho V_{eff}\}$ are helpful (corresponding to Model3 above). In the following, we therefore present the results of Gaussian process regression using this variables set. We experimented with fewer variables (corresponding to Model1 and Model2 above) and obtained worse results.

The results with GPR fits to weighted and unweighted data and with and without the $\rho V_{eff}$ term are also given in **Table 1**. The numbers in the table are for a committee of 5 GPR (to avoid bias associated with a specific draw of train points, as explained in section 2.2). However, the use of a committee leads to an only minor drop (on the order of 10-20%) in *rmse* of the committee model vs the average *rmse* over individual committee members, signifying a largely non-random nature of the errors; for this reason, it is meaningful to report the committee results only. The correlation plots corresponding to these models are shown in **Figure 11** and **Figure 12** for unweighted and weighted data selection, respectively. As expected, the use of more data improves the quality of the fit; with 5,000 training points, very high values of $R^2$ (approaching 0.9999) and very low *rmse* ($1\times10^{-4}$ a.u. or even better) can be obtained.

What is strikingly different in GPR compared to the linear regressions is the role played by the $\rho V_{eff}$ term: while it improved the correlation coefficients and *rmse* but not *B'* in linear fits, it drastically improved and the fit error and *B'* (i.e. quality of energy prediction) with GPR. Very high $R^2$ coefficients and very low *rmse* values are obtained when including $\rho V_{eff}$. Without the $\rho V_{eff}$ term, $R^2$ coefficients and *rmse* values are still better than those obtained with linear regressions, but the model cannot predict positive *B'* values for all three materials simultaneously. With the $\rho V_{eff}$ term, the GPR model is able to predict decent *B'* values for all three materials simultaneously. With GPR (when including $\rho V_{eff}$) we obtained a somewhat better quality of the energy-volume dependence than with best linear regressions.

Data weighting appears to be less impactful with GPR than with linear regressions; it does not fix the problem of negative *B'* values without the $\rho V_{eff}$ term and does not improve the quality of the energy-volume dependence with it. Data weighting is done qualitatively differently in linear and Gaussian process regressions: while in linear regression it helps diminish the weight of values which are in large numbers which, without weighting, prevent the linear model from fitting the tails of the distribution, in the case of GPR weighted data selection leads to uneven sampling of the space of density-dependent variables which may not be advantageous and may be detrimental to model quality.

## 4  Conclusions

We compared the quality of kinetic energy densities and of kinetic energies, specifically energy-volume dependences, following from several known approximations (Thomas-Fermi, von Weiszacker, 4$^{th}$ order gradient expansion (GE4) models as well as a couple recent Laplacian-level KEDFs), linear regression on different sets of density dependent variables, and Gaussian process regressions for two lights metals (Al and Mg) and one semiconductor (Si). The TF, vW, TF+vW, GE4, as well as LKT and PGSL approximations did not show good correlation with the (positive-definite) Kohn-Sham KED but (with the exception of vW) could provide reasonable energy-volume trends, specifically, providing minima near the equilibrium geometry for all three materials simultaneously. Among these models, the 4$^{th}$ order gradient expansion was the best in terms of the quality of the energy-volume dependence.

We achieved good quality KED with linear models which were also able to reproduce well the energy-volume dependence. The best linear fits utilized all terms of the 4$^{th}$ order gradient expansion and weighted fits with weights based on the KED histogram. These regressions resulted in a quality of KED and of the energy exceeding that of any of TF, vW, TF+vW, GE4, LKT, and PGSL approximations. We were able to perform linear fits of the KED of a quality approaching that achieved with the generally more powerful Gaussian process regression. This is not surprising in the sense that, as is well known from neural network literature, it is possible to form sets of coordinates such that a multivariate function can be represented with any accuracy by their linear combination. In other words, non-linear regression, including common ML methods, can always be reformulated as linear regression in new variables, however many. The Taylor expansion is in fact just that – a linear fit with non-linear variables. Using terms of the gradient expansion as variables in a regression model is therefore a natural choice. In this optic, non-surprisingly, we find inclusion of the 4$^{th}$ order term quite useful. This is in line with recent results, e.g. the PGSL functional. Linear regressions are robust and computationally cheap, and our results indicate that development of sets of density-dependent variables suitable for high-quality KED linear regressions is a viable route of development of KEFs.

The variables set which appeared to be the best in linear regressions was also found to be the best in Gaussian process regressions. This suggests a practically useful approach of variable selection based on linear regressions which are relatively easy to perform, and then using those variables with more involved methods such as machine learning methods or other types of non-linear regressions. With GPR, we were able to obtain a much better quality of the KED than with linear models, but the quality of the energy was not much better than with the best linear models. We found that adding $\rho V_{eff}$ as a density-dependent variable substantially improved the quality of GPR, both of the KED and of the energy-volume dependence. In contrast, this term improved the KED but not the energy in linear regressions.

GPR used here used very small train sets (barely a few % of the total data) which indicates their high predictive power. Using even more training data is likely to improve the quality of KED prediction further; however, GPR is a costly method in both training and recall as it involves using the inverse of a matrix of size $N_{train} \times N_{train}$. While 5,000 training points used in this work made the training feasible on a personal computer, using more than 10,000 training points would necessitate high-performance computing resources.

We highlighted the extremely uneven nature of distributions of the KED and of density-dependent variables. We explored weighted regression and weighted data selection based on the histogram of data distribution. We found it to be highly useful in linear regressions but not so much in GPR. This likely has to do with under-sampling of regions in the space of density-dependent variables when used with GPR. Histogram-based weighting is general and will likely be useful in future works on KE(D)F development and in other applications dealing with highly unevenly distributed data, in particular in the context of machine-learning.

## 5 Acknowledgements

This research was enabled in part by support provided by WestGrid (www.westgrid.ca) and Compute Canada (www.computecanada.ca).

## 6 Data availability statement

The data that support the findings of this study are available from the corresponding author upon reasonable request.

## 8    Figures

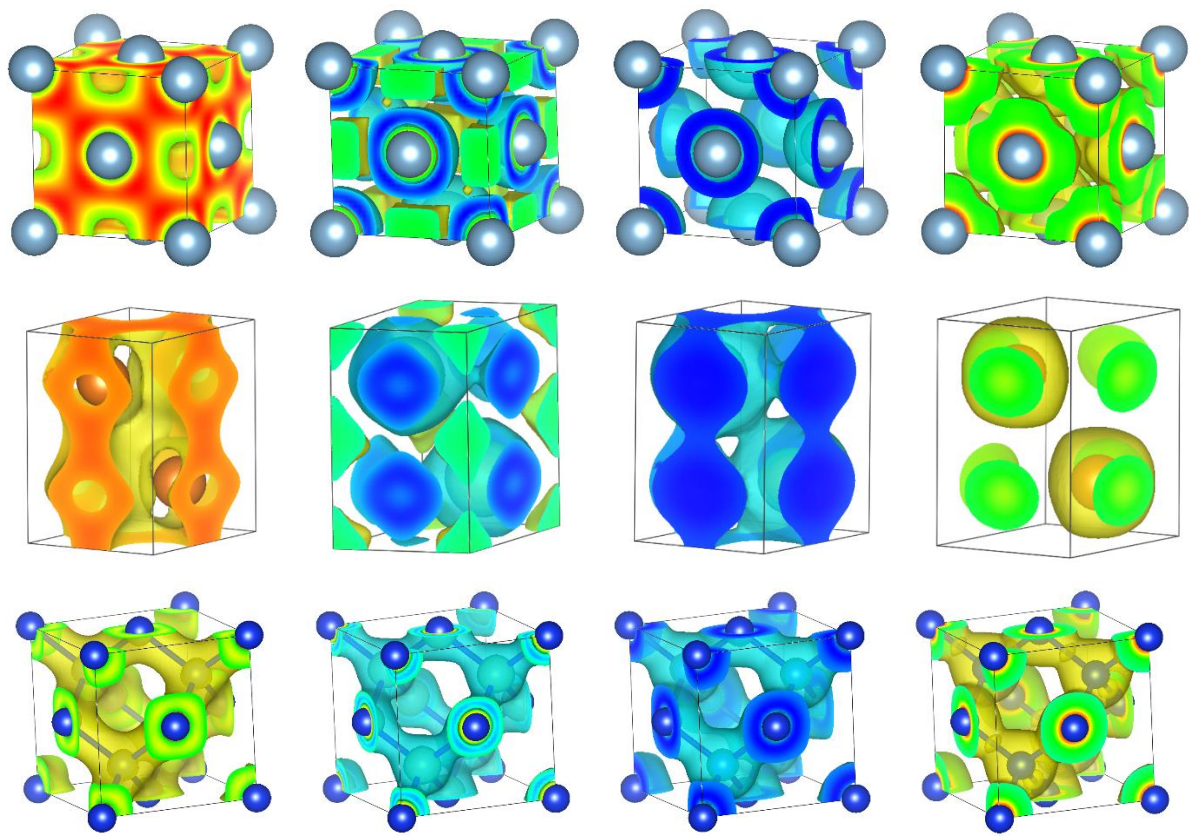

**Figure 1**. Left to right: top row: isovalues of electron density (at 0.02 *a.u.*), Laplacian of the density (at 0.01 *a.u.*), the KS effective potential (at 0.5 *a.u.*), and the KED (at 0.007 *a.u.*) of a unit cell of Al at equilibrium geometry. Middle row: isovalues of electron density (at 0.0135 *a.u.*), Laplacian of the density (at 0.003 *a.u.*), the KS effective potential (at 0.3 *a.u.*), and the KED (at 0.0021 *a.u.*) of a unit cell of Mg at equilibrium geometry. Bottom row: isovalues of electron density (at 0.04 *a.u.*), Laplacian of the density (at 0.03 *a.u.*), the KS effective potential (at 0.4 *a.u.*), and the KED (at 0.013 *a.u.*) of a unit cell of Si at equilibrium geometry. Visualization by VESTA.[59]

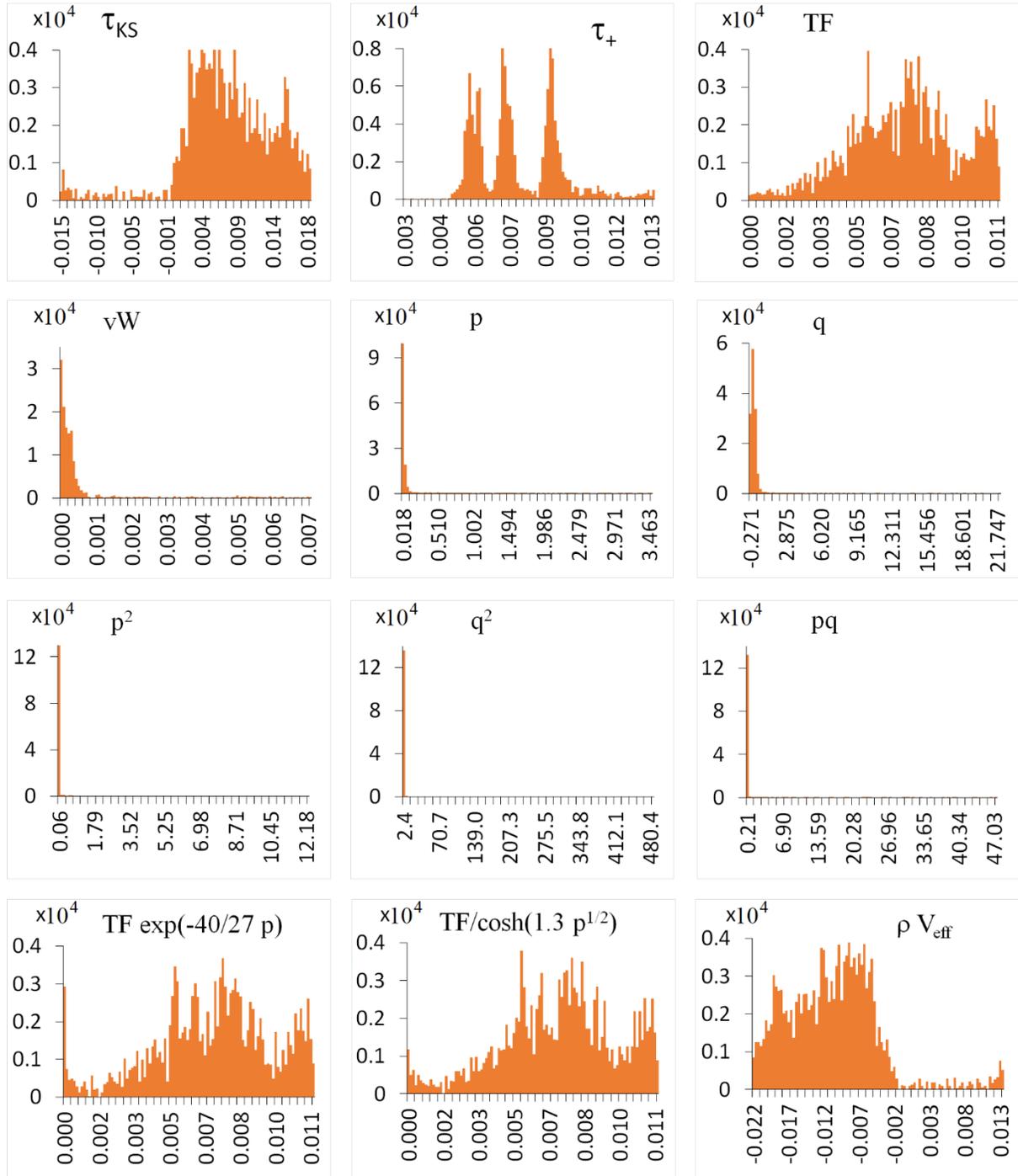

**Figure 2**. Distributions (histograms with 100 bins) of KED and density-dependent variables studied here, for Al at three geometries (equilibrium, compressed, and expanded by 5%). The *x* axis are values of corresponding density-dependent quantities and the *y* axis is the number of values.

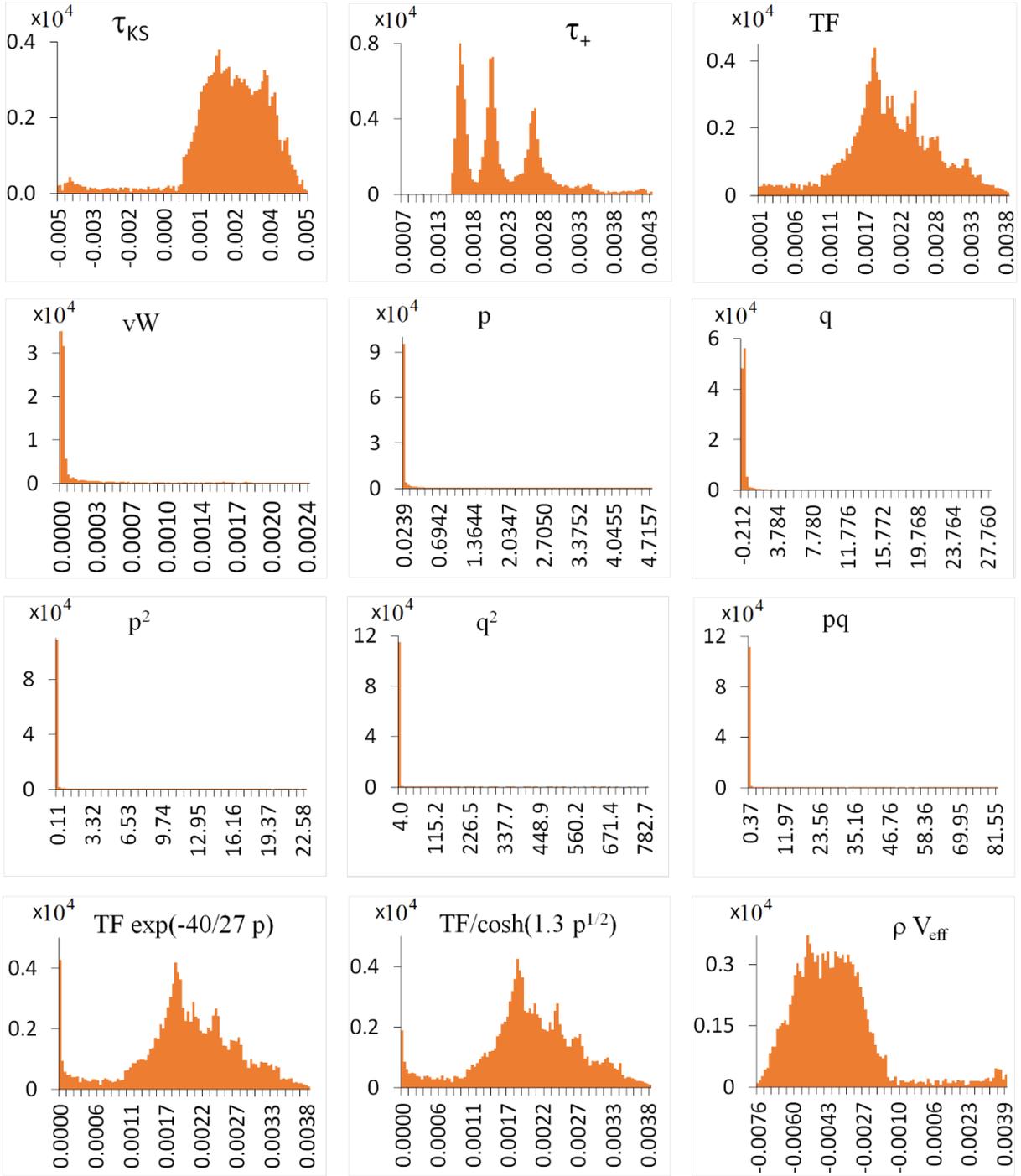

**Figure 3**. Distributions (histograms with 100 bins) of KED and density-dependent variables studied here, for Mg at three geometries (equilibrium, compressed, and expanded by 5%). The *x* axis are values of corresponding density-dependent quantities and the *y* axis is the number of values.

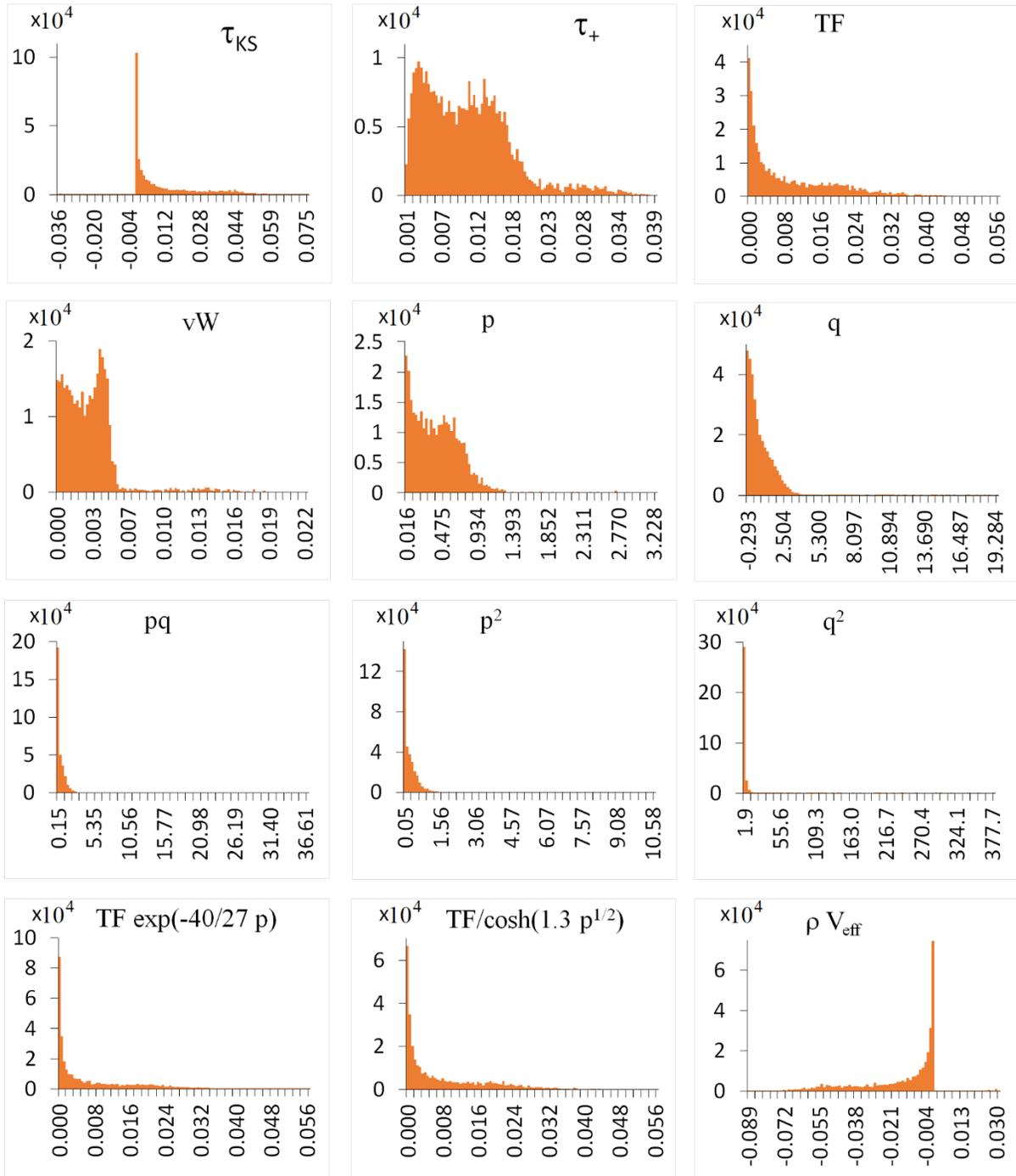

**Figure 4.** Distributions (histograms with 100 bins) of KED and density-dependent variables studied here, for Si at three geometries (equilibrium, compressed, and expanded by 5%). The *x* axis are values of corresponding density-dependent quantities and the *y* axis is the number of values.

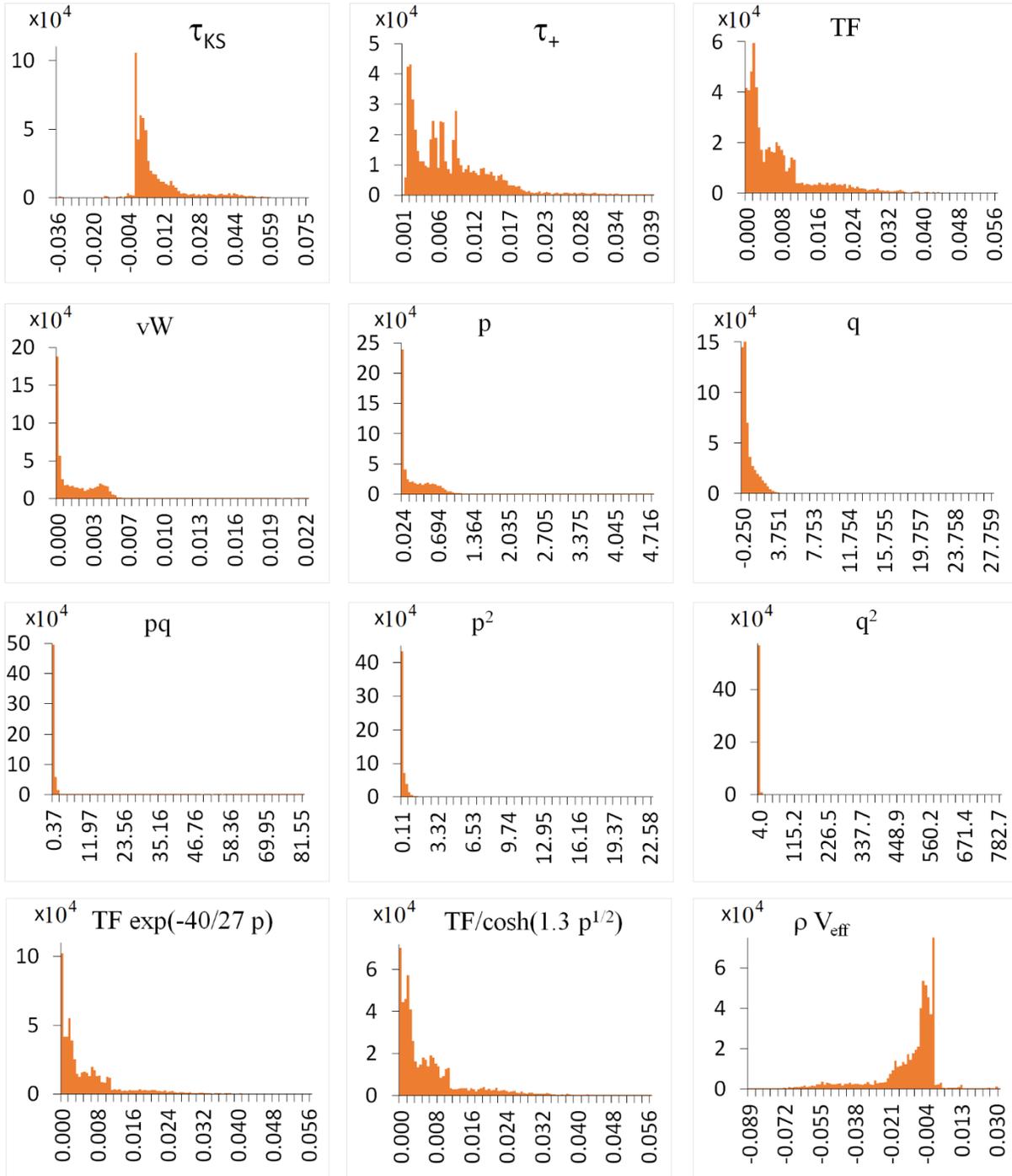

**Figure 5**. Distributions (histograms with 100 bins) of KED and density-dependent variables studied here, combined for Al, Mg, and Si at three geometries (equilibrium, compressed, and expanded by 5%). The *x* axis are values of corresponding density-dependent quantities and the *y* axis is the number of values.

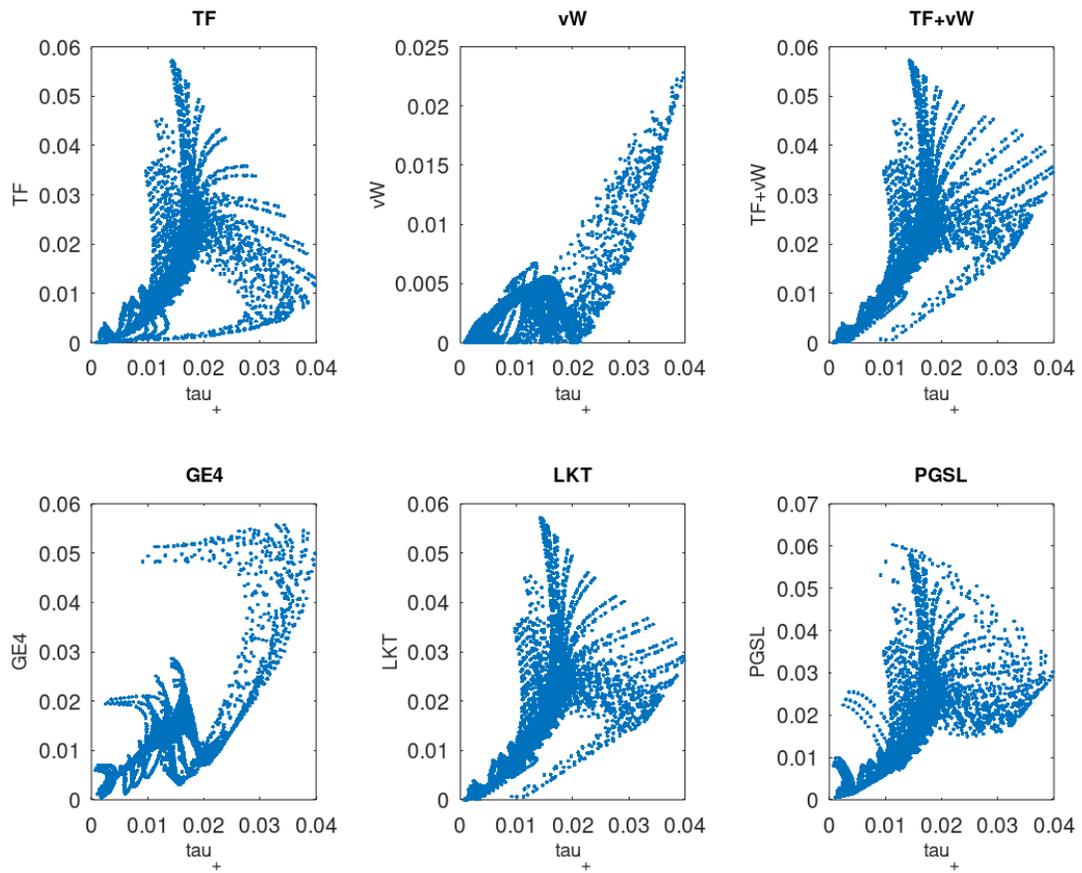

**Figure 6.** Correlations between the KS KED ($\tau_+$) and selected known KEDFs.

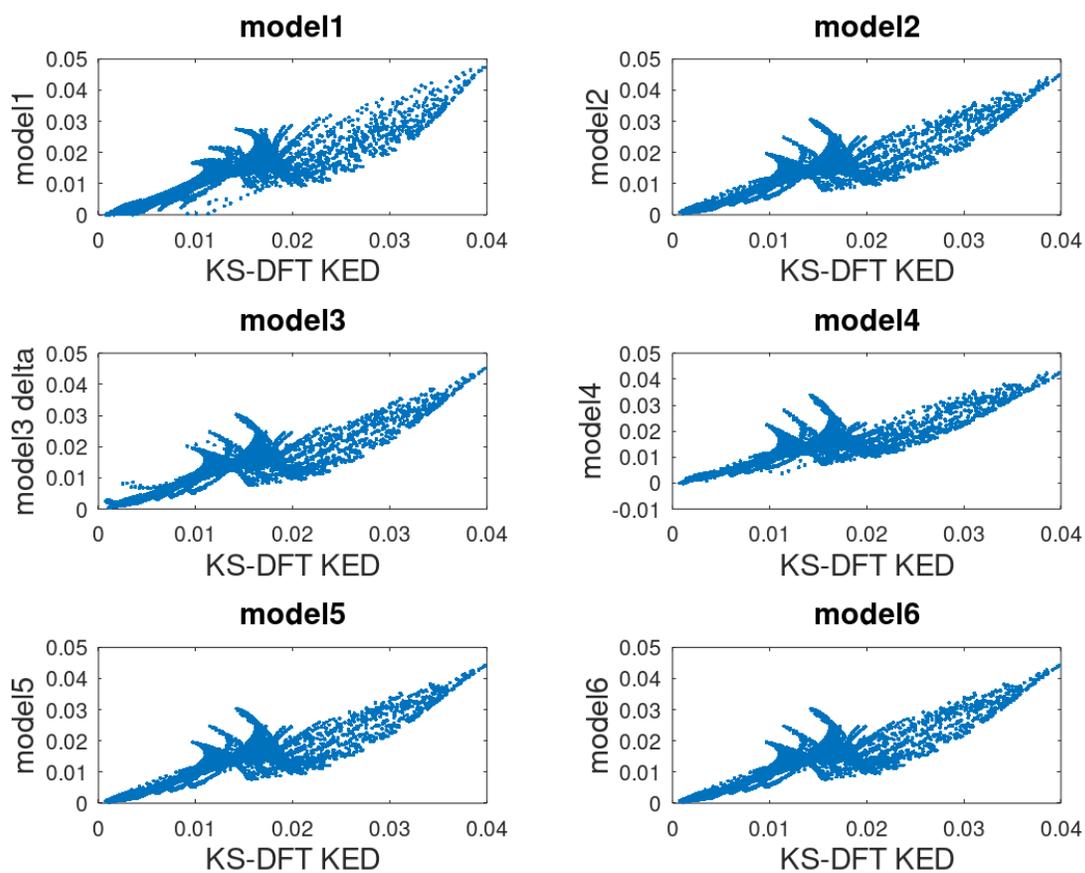

**Figure 7**. Correlation plots between selected KED models and KS KED following from *un*weighted regressions ($\rho V_{eff}$ term not included).

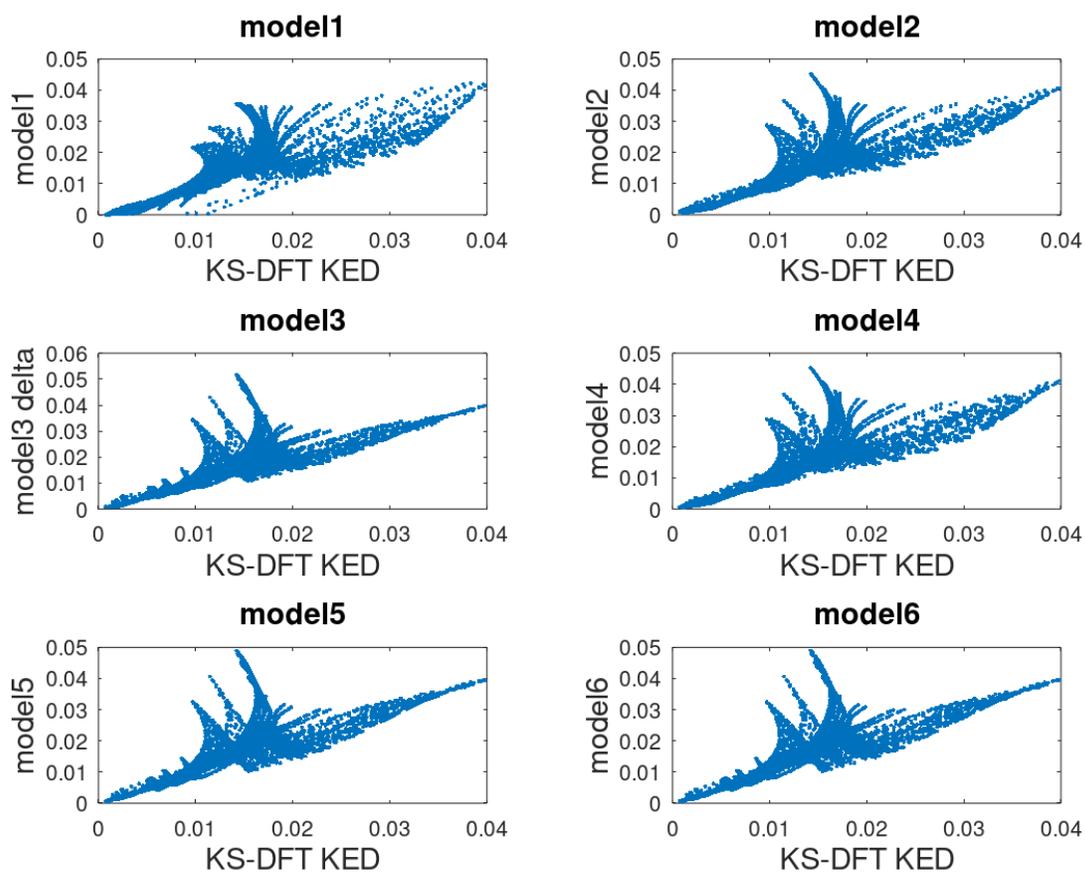

**Figure 8.** Correlation plots between selected KED models and KS KED following from histogram-weighted regressions ($\rho V_{eff}$ term not included).

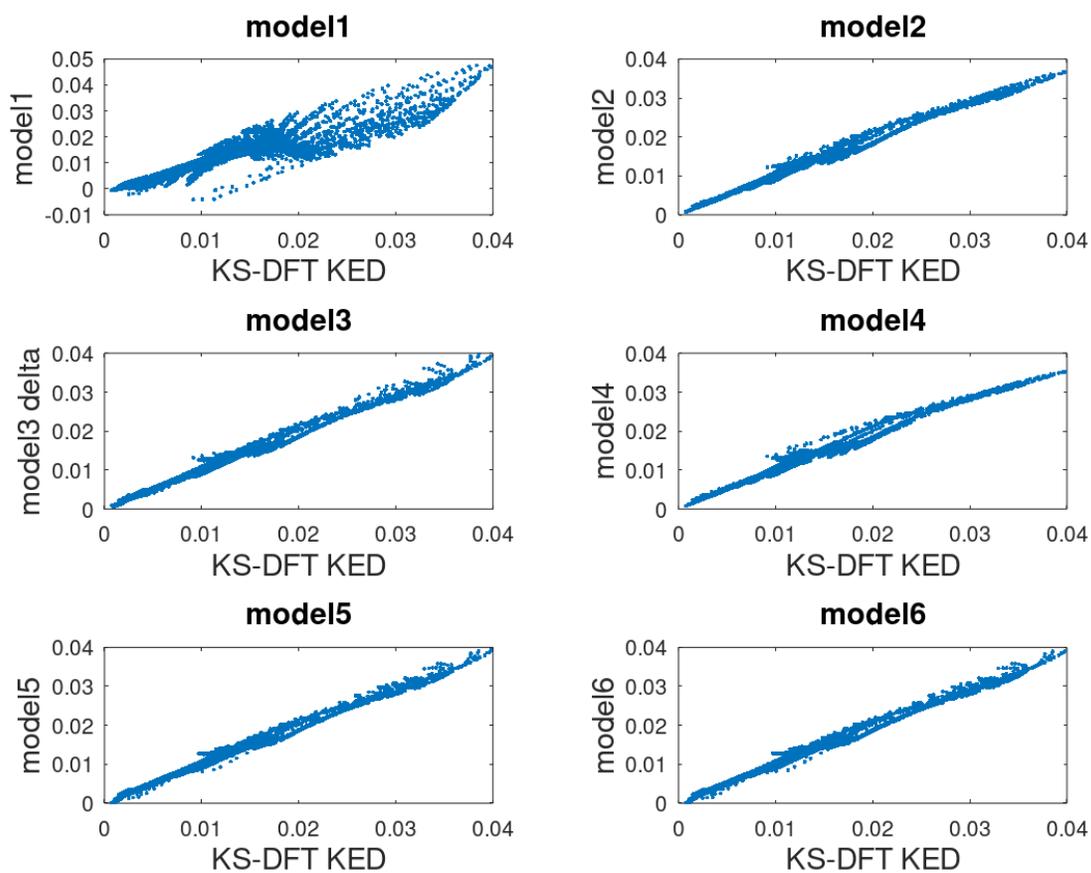

**Figure 9.** Correlation plots between selected KED models and KS KED following from *un*weighted regressions ($\rho V_{eff}$ term included).

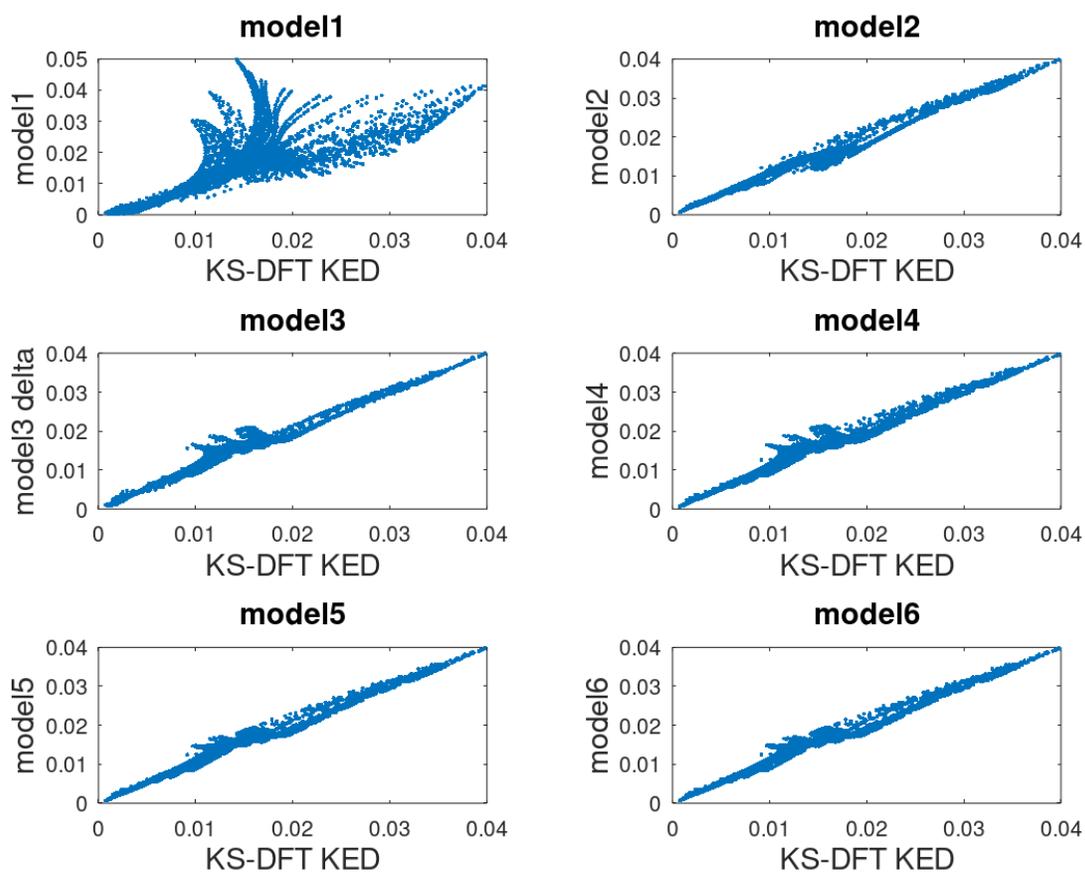

**Figure 10.** Correlation plots between selected KED models and KS KED following from weighted regressions ($\rho V_{eff}$ term included).

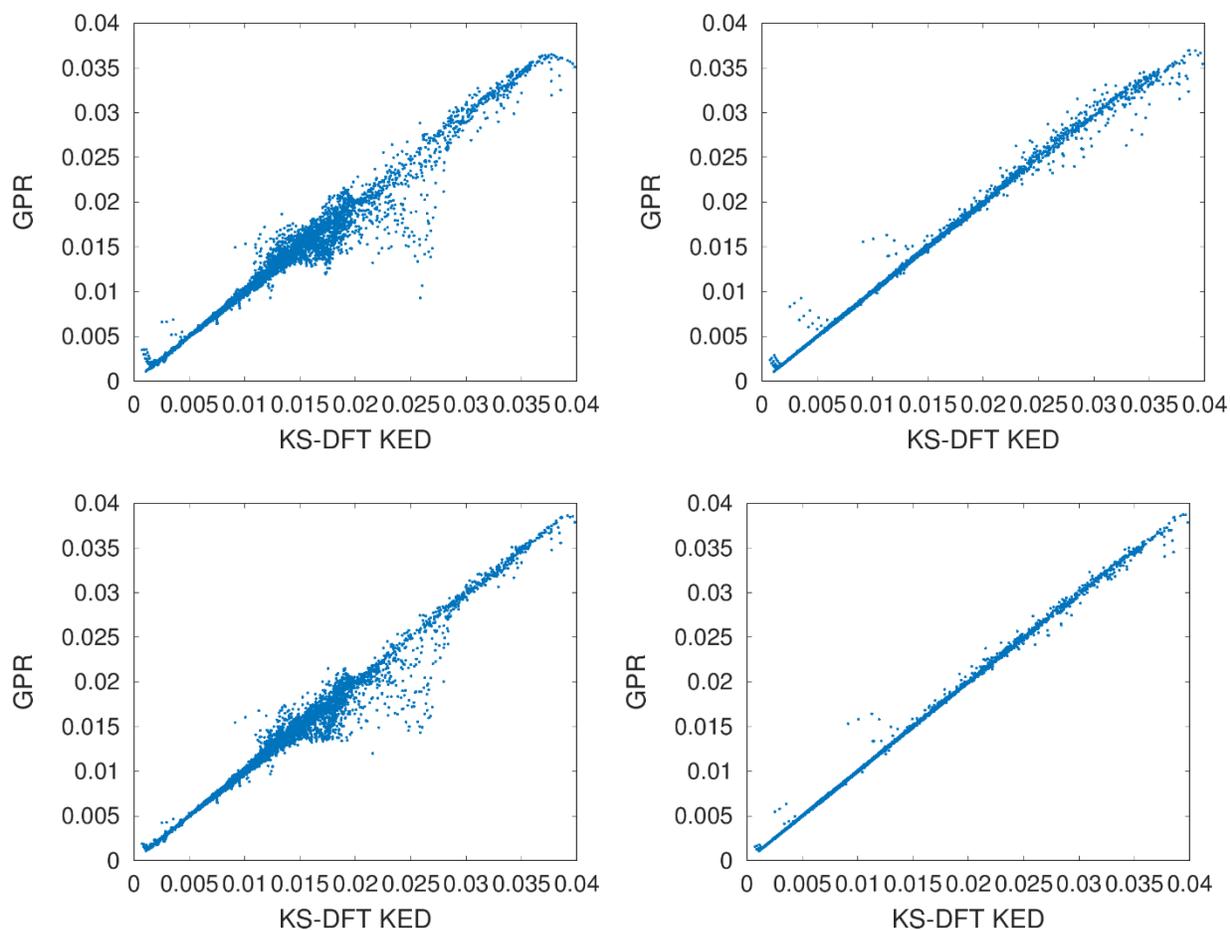

**Figure 11**. Correlation plots between selected GPR models (a committee of 5 GPR) and KS KED following from *un*weighted regressions. Left column: $\rho V_{eff}$ term not included, right column: $\rho V_{eff}$ term included. Top row: 2,000 training points per GPR, bottom row: 5,000 training points per GPR.

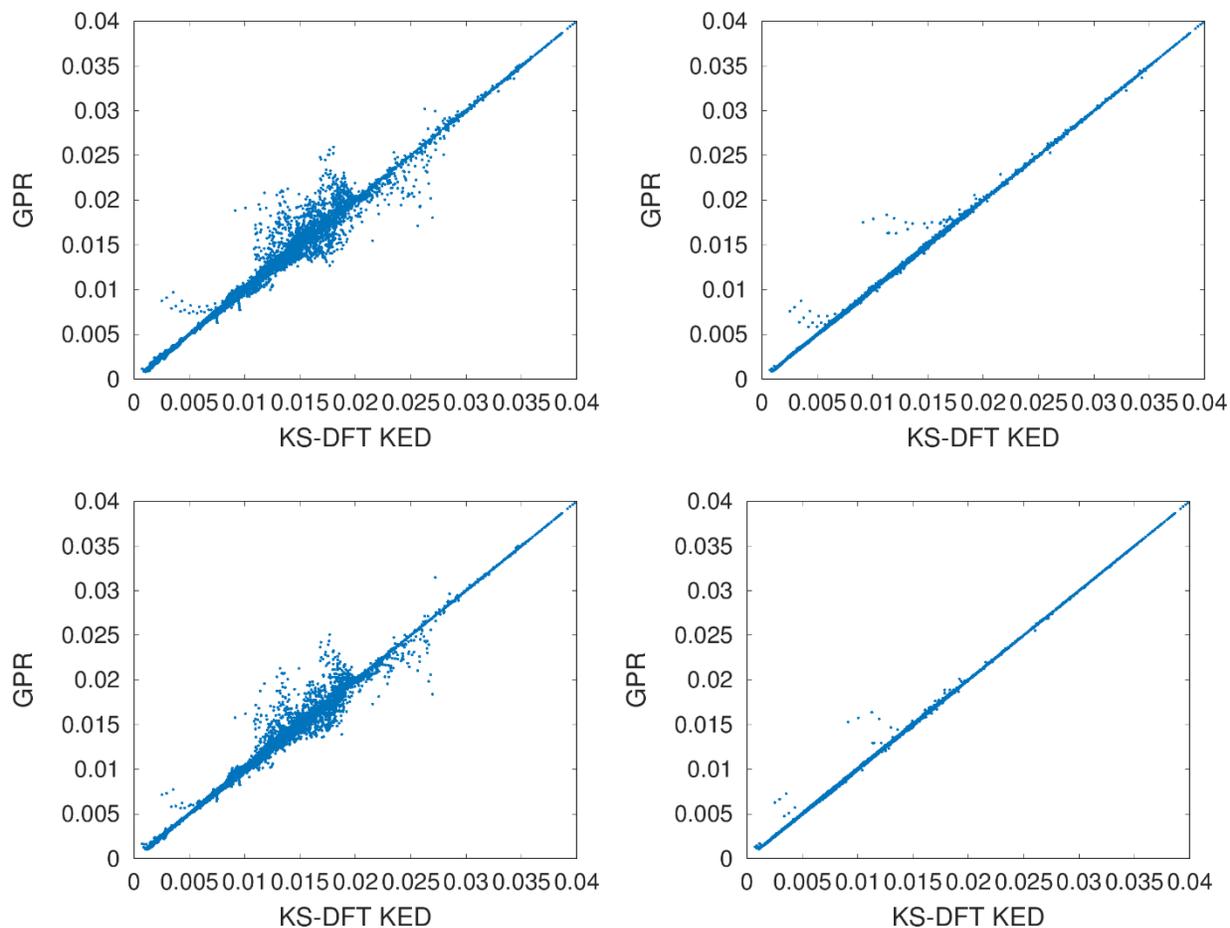

**Figure 12**. Correlation plots between selected GPR models (a committee of 5 GPR) and KS KED following from weighted regressions. Left column: $\rho V_{eff}$ term not included, right column: $\rho V_{eff}$ term included. Top row: 2,000 training points per GPR, bottom row: 5,000 training points per GPR.